\documentclass[iop]{emulateapj}
\usepackage{graphicx,amsmath,url}
\usepackage{natbib}
\usepackage{bm}
\usepackage{hyperref}
\usepackage{xfrac}

\newcommand{\B}{{\bm B}}

\newcommand{\J}{{\bm J}}

\newcommand{\barv}{{\bm{\bar{v}}}}
\newcommand{\er}{{\bm{\hat{e}}_r}}
\newcommand{\ephi}{{\bm{\hat{e}}_\phi}}
\newcommand{\ex}{{\bm{\hat{e}}_x}}
\newcommand{\ey}{{\bm{\hat{e}}_y}}
\newcommand{\ez}{{\bm{\hat{e}}_z}}
\newcommand{\FFF}{\mbox{ ${\cal F}$} {}}
\newcommand{\AAA}{\mbox{ ${\cal A}$} {}}
\newcommand{\BBB}{\mbox{ ${\cal B}$} {}}
\newcommand{\CCC}{\mbox{ ${\cal C}$} {}}
\newcommand{\DDD}{\mbox{ ${\cal D}$} {}}
\newcommand{\EEE}{\mbox{ ${\cal E}$} {}}
\newcommand{\QQQ}{\mbox{ ${\cal Q}$} {}}

\newcommand{\FE}{\FFF_E}
\newcommand{\nab}{\mbox{\boldmath $\nabla$} {}}
\newcommand{\BB}{\mbox{\boldmath $B$} {}}

\newcommand{\EE}{\mbox{\boldmath $E$} {}}
\newcommand{\JJ}{\mbox{\boldmath $J$} {}}
\newcommand{\FF}{\mbox{\boldmath $F$} {}}
\newcommand{\vv}{\mbox{\boldmath $v$} {}}

\newcommand{\uu}{\mbox{\boldmath $u$} {}}

\newcommand{\SSS}{\mbox{\boldmath $S$} {}}
\newcommand{\ov}{\overline}

\renewcommand{\v}[1]{{\boldsymbol{#1}}}

\newcommand{\Eq}[1]{Eq.~(\ref{#1})}

\newcommand{\eq}[1]{\Eq{#1}}

\begin{document}

\title{Radial Stresses and Energy Transport in Accretion Disks}
\shorttitle{Radial Stress in Disks}

\shortauthors{Hubbard et al.}
\author{Alexander~Hubbard\altaffilmark{1},
  Colin~P.~McNally\altaffilmark{2}, Jeffrey
  S. Oishi\altaffilmark{1,3}, \\
Wladimir Lyra\altaffilmark{4,5,6} and Mordecai-Mark Mac Low\altaffilmark{1}}
\altaffiltext{1}{Department of Astrophysics, American Museum of Natural History, New York, NY 10024-5192, USA, {ahubbard@amnh.org,joishi@amnh.org,mordecai@amnh.org}}
\altaffiltext{2}{Niels Bohr International Academy, Niels Bohr
  Institute, Blegdamsvej 17, DK-2100, Copenhagen~\O, Denmark,
  {cmcnally@nbi.dk}}
\altaffiltext{3}{also Department of Physics, Farmingdale State College, 2350 Broadhollow Road, Farmingdale, NY 11735}
\altaffiltext{4}{Jet Propulsion Laboratory, California Institute of Technology, 4800 Oak Grove Drive, Pasadena, CA, 91109, {wlyra@caltech.edu}}
\altaffiltext{5}{Department of Geology and Planetary Sciences, California Institute of Technology, 1200 E. California Ave, Pasadena CA, 91125}
\altaffiltext{6}{NASA Sagan Fellow.}
\begin{abstract}

Early in the study of viscous accretion disks it was
realized that energy transfers from distant sources must
  be important, not least
because the flow at the disk midplane in
the bulk of the disk is likely outwards, out of the gravitational potential well.
If the source of the viscosity is powered by accretion, such as
in the case of the magneto-rotational instability,
such distant energy sources must
lie in the innermost
regions of the disk, where accretion occurs even at the midplane.
We argue
  here that modulations in this energy supply
can alter the 
accretion rate on dynamical, rather than far longer viscous, time scales.
This means that both the steady state value of and fluctuations in the inner disk's accretion rate,
 depending on the details of the inner boundary condition and occurring on the inner disk's rapid
evolution time, can affect the outer disk.
This is particularly interesting because observations have shown that disk accretion is not steady
(e.g.~EX Lupi type objects).
We also note that the power supplied to shearing boxes is 
set by the boxes themselves rather than the physical energy fluxes in a global disk.  That is,
their saturated magnetic field 
is not subject to the full set of energy constraints present in an actual disk.
Our analysis suggests that large scale radial transport of energy 
has a critical impact on the evolution and variability of accretion disks.

\end{abstract}
\keywords{accretion, accretion disks --- instabilities --- magnetic fields --- magnetohydrodynamics --- plasmas}

\section{Introduction}

Accretion disks surround many objects in the universe, from young
stars, to compact binary companions, to supermassive black
holes. Their observed accretion flows  
require outwards transport of angular momentum. The magnetic coupling of regions at
different radii in accretion disks ranks high among the usual suspects
for angular momentum transport mechanisms.  This transport can occur
either radially through the disk, such as with the magnetorotational
instability \citep[MRI,][]{1998RvMP...70....1B}, or vertically through
a disk wind \citep{2000prpl.conf..759K}.  Understanding vertical
angular momentum transport through disk winds depends on
understanding the disk's vertical boundaries.  Similarly,
understanding radial angular momentum transport requires understanding
the radial boundary conditions.  
In this paper, we argue that this is true not only near to but also far from those boundaries. 
While that may seem a trivial statement, 
in practice it means 
that even distant boundary conditions must be taken into account
by numerical simulations intended to study the magnitude and direction
of any accretion flows.

The common
assumption of local control implicitly places any transition
between accretion and decretion far from any region
explicitly considered.  Simulating regions far from any physically
distinct radius further allows the use of the shearing box
approximation, which goes so far as to erase the distinction between
inwards and outwards directions.  \citet[][hereafter BP99]{1999ApJ...521..650B} found
that the mean flow dynamics of MRI-driven MHD turbulence
were a local phenomenon, justifying the use of alpha-disk models and
shearing boxes.

It has long been known that steady state, vertically integrated,
viscous, regions of accretion disks require an outwards energy flux in the standard theory
\citep{1974MNRAS.168..603L}\footnote{See the footnote on page 611 of that article: ``For a Newtonian point mass [...]
the power liberated at radii between $R$ and $R+dR$ is three times larger than the energy generated there.''}.  
That theory defines $r_b$ as the position of the 
boundary layer where the differential rotation
goes to zero \citep{2002apa..book.....F}. 
Protostars rotate below the Keplerian frequency, so $r_b$ must exist for disks that extend to the surface
of their protostar, although it may not exist in strongly magnetized disks \citep{2002apa..book.....F}.
It was shown early in the study of viscous disks that 
dissipation at radii $r > 9r_b/4$
thermalizes significantly more energy than is released by the accretion
flow there 
\citep{1974MNRAS.168..603L, 1998RvMP...70....1B,2002apa..book.....F}, with the balance
made up by an energy flux built up within $ 9r_b/4$, and depleted over
the entire outer disk.

Viscous disk theory predicts that disks
decrete, rather than accrete, in regions where 
the stresses that drive accretion and
angular momentum transport fall off faster with radius than $r^{-2}$
\citep[][BP99]{1974MNRAS.168..603L}.  In standard, self-similar disk models, this decretion condition is
easily satisfied in disk midplanes where the pressure decreases
rapidly with radius.
It takes vertical integration through viscous disks to reliably find net accretion.
This has led to the concept of meridional circulation,
in which a decreting midplane is more than balanced by accreting
surface layers 
\citep{1984SvA....28...50U,1988AcA....38...21S,1992ApJ...397..600K,1994ApJ...423..736R,2002A&A...396..623R,2002ApJ...581.1344T,2004A&A...415.1177K,2007A&A...463..369T,2007Sci...318..613C,2009Icar..200..655C,2010ApJ...719.1633H}. 
Such behavior has been demonstrated in
numerical models with a fixed anomalous viscosity defined by a
constant Shakura-Sunyaev $\alpha$ parameter \citep{1973A&A....24..337S,2011A&A...534A.107F}.

These layered accretion structures arise in uniformly 
viscous disks because
the disk thickness increases with radius in such disks.
While the gas density and pressure in the midplane
decrease with radius, they instead increase with cylindrical 
radius far enough above the midplane 
that the disk surface is encountered at finite radius.  This
occurs at increasing radius with height because the tilted disk
surface reaches higher altitudes at larger radii.
This transition in behavior 
is lost in slab geometries with radially constant scale height such as
the shearing sheet approximation. 
In classical viscous disk theory the stress remains proportional to the gas
pressure, so there exists a dividing height between decretion and accretion that occurs
close to the surface,
where the radial pressure gradient goes from falling off faster to
falling off slower than $r^{-2}$ \citep[e.g.][]{1994ApJ...423..736R}.

Decreting regions of disks must be net importers of energy, since
decretion boosts material out of a potential well.  
These regions still dissipate energy locally through non-ideal effects such as viscosity or resistivity,
just as accretion regions do \citep{1974MNRAS.168..603L},
but that just adds to the energy deficit that must be balanced by
importing energy from inner regions.
However, the source of that energy must be determined by the nature of the stresses driving
accretion and decretion.  If those stresses are powered by the accretion process itself and
occur in the horizontal
$r$-$\phi$ plane, as expected for magnetorational instability
\citep{1991ApJ...376..214B,1998RvMP...70....1B},
We argue that
this means that the midplane outflow of meridional circulation is powered
not by the accreting upper layers, but rather
by the inner edge of the midplane itself.
Similar radial energy flows occur at every altitude, with the
accreting inner edge of the disk at each altitude forming the
accreting upper layer described by the meridional circulation picture.

BP99 argued that the mean flow dynamics of
MRI-driven, MHD turbulence can be treated as a local phenomenon
because the dynamical equations describing the flow are
themselves local:
 the time derivatives at a specific spatial point of quantities such as the
mean velocity or the mean magnetic field
depend only on their values and spatial derivatives at that point.
However, even distant regions of a disk will communicate
 on long enough time scales.
This means that locality is more accurately
treated as a time scale condition: on time scales shorter
than some critical value,
the disk can be treated as local, but it cannot be treated
as local for longer time scales.  

We show that the strength of the radial energy flows required to maintain the disk
  structure determine that critical time scale. 
   The
magnetic fields that provide the Maxwell stresses in MRI active disks have only
a very small energy density, 
comparable to the accretion or decretion power integrated only
over a few orbits. 
Order unity variations in the
energy fluxes 
due to variability in the inner disk 
can therefore, everywhere within a few local orbits, 
cause order unity
variations in the stresses, and hence the local accretion or decretion
rates.
This motivates us to extend the time dependent analysis in BP99
by including the time derivative of the stresses.

In Section~\ref{sec_angmom} we revisit the basic equations for viscous disks.
In Section~\ref{Sec_Fluxes}
we discuss the energy fluxes present and estimate the dynamical time-scale
associated with the stresses.  In Section~\ref{sec:shear-box-appr} we delve more deeply into the energy flux
considerations in the context of the shearing sheet approximation, and discuss the difference
between the case where stresses are powered by accretion energy (which includes
most cases of turbulent viscosity) vs.\ the cases where the stresses are not powered by accretion
energy (including many cases of microphysical viscosity).
In Section~\ref{sec:simul-accr-disks} we discuss links to numerical simulations, both
existing results and suggestions for future design and diagnostics.
In Section~\ref{sec:BP99} we place our results in context as an
extension of the analysis of BP99, and
we conclude in Section~\ref{sec_conclusions}.

\section{Angular Momentum Transport}
\label{sec_angmom}

\subsection{Azimuthal Lorentz Forces}
\label{sec:azim-lorentz-forc}

The action of the magnetic field on the gas to transfer angular
momentum occurs through 
azimuthal Lorentz forces
$\FF_{L,\phi} =\JJ \times \BB|_{\phi}$.  In cylindrical
coordinates, the magnetic field is
\begin{equation}
\BB = B_r \er + B_\phi \ephi + B_z \ez, 
\end{equation}
and the current density 
\begin{align}
\mu_0\JJ =& \nab \times \BB,  \\
 =&\left[ \frac{1}{r} \partial_\phi B_z -\partial_z B_\phi\right] \er \nonumber\\
 & +\left[\frac{\partial B_r}{\partial z} - \frac{\partial B_z}{\partial r} \right] \ephi \nonumber\\ 
 & + \left[\frac{1}{r}\partial_r(r B_\phi) -\frac{1}{r}\partial_\phi B_r \right] \ez.
\end{align}
The  azimuthal component of the Lorentz force 
can be usefully expanded and rearranged:
\begin{align}
\mu_0 F_{L,\phi} =& \mu_0 \left( J_z B_r - J_r B_z\right) \label{eq:lorentz-terms}\\
= & \frac{1}{r}\left[\partial_r (r B_\phi)\right] B_r  - \frac{1}{r} (\partial_\phi B_r) B_r \nonumber  \\
& \qquad  - \frac{1}{r }(\partial_\phi B_z )B_z + (\partial_z B_\phi) B_z\\ 
=& -\frac{1}{2 r} \partial_\phi B_r^2 -\frac{1}{2 r} \partial_\phi B_z^2 +\frac{1}{r^2} \partial_r (r^2B_\phi B_r)  \nonumber\\
& \quad   -\frac{B_\phi}{r} \partial_r (r B_r)  + \partial_z(B_\phi B_z) - B_\phi \partial_z  B_z \\
=& -\frac{1}{2r} \partial_\phi \left( B_r^2 + B_z^2 \right) + \frac{1}{r^2}\partial_r (r^2 B_\phi B_r) \nonumber\\
&  + \partial_z (B_\phi B_z)+B_\phi\left[ -\frac{1}{r} \partial_r (r B_r) - \partial_z B_z \right].  \label{Reduced_Max_0}
\end{align}
We can now invoke  $\nab \cdot \B =0$ to clarify the component terms of the azimuthal Lorentz force:
\begin{align}
\mu_0 F_{L,\phi}&= \underbrace{\frac{1}{r^2} \partial_r (r^2 B_\phi B_r)}_\text{Radial Stress} + \underbrace{\partial_z (B_\phi B_z)}_\text{Vertical Stress} \nonumber \\
 & \qquad \qquad \quad + \underbrace{\frac{1}{2r} \partial_\phi \left( B_\phi^2 - B_r^2 -B_z^2\right)}_\text{Anisotropic magnetic pressure}.
\label{Reduced_Max}
\end{align}
The pressure term 
azimuthally averages to zero due to periodicity \citep[e.g.][]{1973A&A....24..337S, 2003ARA&A..41..555B}.

\subsection{Accretion Stresses}
\label{sec:accretion-stresses}
The components of Equation~(\ref{Reduced_Max}) are derivatives of
the radial and vertical components of the Maxwell stress tensor\footnote{
A common alternative definition for the stress is $\mu_0M_{ij} \equiv B_iB_j -
  \sfrac{1}{2}\,\delta_{ij} B^2$, but this only matters for the diagonal (pressure) components.} 
\begin{equation}
M_{ij} \equiv \mu_0^{-1}B_i B_j, 
\end{equation}
The evolution of angular momentum in disks
includes contributions from both the magnetic Maxwell stress
and the hydrodynamical Reynolds stress 
\begin{equation}
R_{ij} \equiv \rho v_i v_j, 
\end{equation}
often combined in a total stress tensor
\begin{equation}
T_{ij}  \equiv R_{ij} - M_{ij}.\label{eq_Tdef}
\end{equation}
We emphasize here that in the above equations we use the full, not fluctuating,
velocities and magnetic fields.
We can write the time evolution of the angular momentum density in terms of the stresses as
\begin{equation}
\partial_t \langle \rho r v_\phi \rangle_\phi  + \frac
1r \partial_r  \langle r^2 T_{\phi r}\rangle_\phi + r \partial_z  \langle T_{\phi
  z}\rangle_\phi=0 \label{stress_0}.
\end{equation}
where we have denoted averaging performed over the dimension $i$ with
the notation $\langle \dots \rangle_{i}$. 
The azimuthal averaging here again eliminates the azimuthal pressure term.
(Equation~(\ref{stress_0}) is 
often vertically averaged as well.)
We name stress terms such as $M_{\phi r}$
``horizontal'' while terms such as $M_{\phi z}$ are labelled
``vertical''. 

\subsubsection{Alternate Definitions of the Stresses}
\label{sec_rey_decomp}

It is common to decompose the velocities and magnetic field into mean (generally azimuthal
averages) and fluctuating terms, 
marked here with overbars and primes respectively, while taking the 
azimuthal density fluctuations to vanish.  
In that case, Equation~(\ref{eq_Tdef}) becomes
\begin{equation}
\langle T_{ij}\rangle_\phi  = \rho \ov{v}_i\ov{v}_j - \mu_0^{-1} \ov{B}_i \ov{B}_j+\langle\rho v_i' v_j' \rangle_\phi -
\mu_0^{-1} \langle B_i'B_j'\rangle_\phi.\label{eq_Tdef2}
\end{equation}

When moving from Equation~(\ref{eq_Tdef2}) to
Equation~(\ref{stress_0}), two terms are often treated differently: $\rho \ov{v}_i \ov{v}_j$ 
is often separated from the stress, and written as the radial
advection of angular momentum; while 
the term $\mu_0^{-1} \ov{B}_i\ov{B}_j$
is often neglected in determining the 
\citeauthor{1973A&A....24..337S}~$\alpha$ parameter. 

\subsubsection{Independence of Torques from Density Gradients}

Past studies, such as BP99, have chosen to perform a density weighted
vertical and azimuthal average of Equation~(\ref{stress_0}).  The
horizontal stress term is given by 
\begin{equation}
\int dz \left\langle \frac 1r \partial_r \left(r^2 T_{\phi r}\right)\right\rangle_{\triangle r,\phi}= \frac 1r \partial_r \left( r^2 \varSigma W_{r\phi}\right)
 \label{stress_1},
\end{equation}
where $\triangle r$ represents a small interval in $r$, 
the disk surface density $\varSigma$ is given by
\begin{equation} \varSigma = \int dz \langle \rho
  \rangle_{\triangle r,\phi}, \end{equation}
and
\begin{equation}
W_{r \phi} \equiv \varSigma^{-1} \int dz \langle  T_{\phi r}
\rangle_{\triangle r, \phi} \label{W_def}.
\end{equation}
For a truly viscous, azimuthally symmetric disk, the $T_{\phi r}$ term in Equation~(\ref{stress_0}) is:
\begin{equation}
\int dz\left(\frac1r \partial_r \langle r^2 T_{\phi r}\rangle_{\triangle r\phi}\right)=-\frac 1r \partial_r \left( r^3 \nu \varSigma \partial_r \varOmega \right),
\end{equation}
where $\nu$ is the kinematic viscosity \citep{1974MNRAS.168..603L}.
This averaging serves the purpose of showing that the horizontal Maxwell and Reynolds
stresses act like a viscosity if we equate
\begin{equation}
W_{r \phi} \sim -\nu r \partial_r \varOmega,
\end{equation}
and induces no mathematical error. However, it can lead to confusion: a naive reading of Equation~(\ref{stress_1}) suggests that a spatially varying surface density
$\varSigma$ can create torques from a Maxwell stress with radially constant $r^2 M_{\phi r}$, in contradiction to
Equation~(\ref{stress_0}).  Equation~(\ref{W_def}) makes clear,
however, that $W_{\phi r}$ is defined by 
dividing by $\varSigma$, so $W_{\phi r}$ and $\varSigma$ cannot be varied independently.

\subsection{Condition for Accretion}
\label{subsec:direction}

Equation~(\ref{Reduced_Max}) tells us that the effect of the azimuthal force deriving from $M_{\phi r}$
 depends on the sign of the product of
$\partial_r (r^2 B_\phi B_r)$ with the
angular velocity $\varOmega$. If 
\begin{equation} 
 \varOmega \partial_r (r^2 B_\phi B_r) > 0,
\end{equation} 
then the Lorentz force exerts a
torque on the disk aligned with the orbital rotation, resulting in
decretion, while if it is negative, the Lorentz 
force exerts a torque acting against the rotation, resulting in an accretion flow.

Driving an
accretion flow then requires
\begin{equation}
\label{Criterion}
\partial_r \left[r^2 (-M_{\phi r}) \right]>0
\end{equation}
(from Eq.~\ref{Reduced_Max}).
This can be rephrased as the statement that accretion requires that
the magnitude of $M_{\phi r}$
have a radial dependence shallower than $r^{-2}$.

\subsection{Stresses Proportional to Pressure}
\label{sec_T_propto_p}

If the only physical quantities of the background disk are its density $\rho$ and sound speed $c_s$, then the
the stresses must scale with the thermal pressure, the only combination of the parameters with the correct dimensions:
$\rho_0 c_s^2$ (the standard case of an $\alpha$ disk, \citealt{1973A&A....24..337S}).  This means that
\begin{equation}
M_{\phi r} \propto \rho_0 c_s^2 \propto H \varSigma  r^{-3}, \label{ZNF_scaling}
\end{equation}
where we have used $\varOmega^2 \propto r^{-3}$.  For such a region 
of a Keplerian disk to accrete,  
the radial dependence of $H \varSigma$ must exceed
$r^{+1}$.
Observations suggest that the surface density dependence for real accretion
disks cannot be much shallower than $\varSigma \propto r^{-1}$ 
\citep{2003MNRAS.341..501S,2007prpl.conf..555D,2010ApJ...723.1241A}, 
so to produce accretion the minimum disk flaring must be $H\propto r^2$.
The standard disk model with constant opening angle and $\varSigma \propto r^{-1}$ 
has $H\propto r$, so it must have horizontal Maxwell stresses 
that drive decretion instead of accretion.

Of course, Equation~(\ref{ZNF_scaling}) is local, and the scaling applies only to the midplane.  Vertically integrating it
leads to 
\begin{equation}
\int_z dz M_{\phi r} \propto H^2 \varSigma r^{-3},
\end{equation}
which will generally fall off slower than $r^{-2}$, recovering the net accretion flow of viscous disk theory.  Such a scenario
is known 
\citep{2002ApJ...581.1344T}
to have a meridional circulation with a decreting midplane as above counterbalanced by
accreting surface layers. 

\subsubsection{Stresses Not Proportional to Pressure}

Especially in the case of magnetized accretion flows, more physics is expected, which provides
additional parameters that can control the stresses.
For example, the stresses in the MRI case likely depend on any net background vertical
magnetic field, which has no reason to be proportional to the midplane pressure, 
and the stresses
can certainly depend on non-ideal effects.  
Further, in the MRI case any estimation $M_{\phi r} \propto \rho c_s^2$ 
applies only in regions where the magnetic fields are locally generated,
and simulations have shown that the upper layers of accretion are
actually coronas powered by magnetic field generated closer to the midplane
\citep[e.g.][]{2000ApJ...534..398M,2006A&A...457..343F,2009ApJ...704L.113B,2010MNRAS.405...41G,2010ApJ...708.1716S,2011ApJ...735..122F,2014MNRAS.438.2513P}.  
This means that $M_{\phi r}$ will not be proportional to the thermal pressure precisely in the upper layers where
standard meridional circulation models predict accretion flows, which may help explain why
\cite{2011A&A...534A.107F}'s MRI simulations saw both decreting midplanes and surface layers.
Nonetheless, in the absence of dominant imposed physics, standard
disk radial scalings imply decreting midplanes.

\section{Energy flows}
\label{Sec_Fluxes}

It is puzzling that decretion flows are predicted, and seen in numerical simulations.
Where is the origin of the energy
required to drive such a flow 
out of a gravitational potential? 
If a volume embedded in the disk is locally accreting, then the local loss of gravitational potential
energy can provide the energy dissipated during turbulent angular momentum transport.
If it is locally decreting, though, then some distant energy
source must provide the energy for both the decretion flow
and the associated turbulent dissipation.

\subsection{Energy Transport by Poynting Flux}
\label{sec_E_transport}

Radial fluxes of energy (in the magnetic case, Poynting fluxes) are
known to be dynamically significant
in disks accreting through horizontal viscous stresses
\citep{1974MNRAS.168..603L, 1998RvMP...70....1B,2002apa..book.....F}.
We can calculate the time evolution of the magnetic energy density:
\begin{equation}
\partial_t \frac{B^2}{2 \mu_0} = \BB \cdot \left[\nab \times \left( -\EE\right)\right], \label{E0}
\end{equation}
where the electric field $\EE= -\vv \times \BB+ \eta \JJ$,
with $\eta$ being the resistivity.

Noting that
\begin{align}
\BB \cdot \left[\nab \times \left( -\EE\right)\right] &= -\EE \cdot \left(\nab \times \BB\right) -\nab \cdot \left( \EE \times \BB\right)  \\
&=\left(\vv \times \BB\right) \cdot \JJ  -\eta \JJ^2 -\nab \cdot \SSS,
\end{align}
where $\SSS = \mu_0^{-1} \EE \times \BB$ is the Poynting flux, we can rewrite Equation~(\ref{E0}) as
\begin{equation}
\partial_t \frac{B^2}{2 \mu_0} = -\nab \cdot \SSS - \eta \JJ^2 - \FF_L \cdot \vv, \label{E1}
\end{equation}
where $\FF_L \equiv \JJ \times \BB$ is 
the Lorentz force.  Equation~(\ref{E1})
states that, up to resistive terms, any energy taken from the kinetic flow
by the Lorentz force ($-\FF_L \cdot \vv$)
goes into the magnetic field (either locally, or, through the Poynting
flux, elsewhere) and vice-versa.
In a decretion flow azimuthal forces are aligned with the orbital motion, torquing up the disk.  Therefore in
Equation~(\ref{E1}) decretion flows imply $\FF_L \cdot \vv>0$.
To balance Equation~(\ref{E1}) the power required to torque up the disk must come from either
the local magnetic energy through the term $\partial_t B^2$, or from a
deposit of magnetic energy generated elsewhere and transported by the Poynting flux ($\nab \cdot \SSS$).

Calculating the Poynting flux using only the orbital velocity $\vv = r \varOmega \ephi$ we find
\begin{align}
\SSS =& \left(-\vv \times \BB\right)\times \BB/\mu_0   \label{SS_Orb} \\
&= -r \varOmega M_{\phi r} \er - r \varOmega M_{\phi z} \ez + r \varOmega \left(B_r^2 + B_z^2\right) \ephi. \nonumber
\end{align}
Taking the divergence of Equation~(\ref{SS_Orb}) and averaging in azimuth we find
\begin{equation}
\langle \nab \cdot \SSS \rangle_{\phi} = -\frac 1r \partial_r \left(r^2 \varOmega \langle M_{\phi r}\rangle_\phi\right) - \partial_z \left(r \varOmega
M_{\phi z}\rangle_\phi\right). \label{divS_Orb}
\end{equation}
Equation~(\ref{SS_Orb})
shows that horizontal stresses (such as $M_{\phi r}$)
are associated with radial energy transport, while vertical stresses
are associated with vertical energy transport.
In disks with meridional circulation  accreting through horizontal stresses, energy released in surface
layers at a radius $r$ therefore does not travel vertically
to power the decreting midplane at that same radius.

The horizontal component of Equation~(\ref{divS_Orb})
can be approximated by assuming a power-law behavior 
\begin{equation}
M_{\phi r} \propto r^{-s}
\end{equation}
to find
\begin{equation}
\langle \nab \cdot \SSS \rangle_{\phi} =(1/2-s )  \varOmega  \langle M_{\phi r} \rangle_{\phi}.
\label{Poy1}
\end{equation}
For comparison to this value, the local power released by accretion or required by
decretion is approximately 
\begin{equation}
 \frac 32 \varOmega  \langle M_{\phi r} \rangle_{\phi,t}.
 \label{eq:poy2}
\end{equation}
Because $s$ is of order unity, \eq{Poy1} and \eq{eq:poy2} are of the same order, except in the 
special case of $s=1/2$. 
Thus, the Poynting flux extracts an 
order unity fraction of the accretion power from accreting regions,
and injects similar amounts in decretion regions (again, except in the
special case of $s=1/2$). 

While we focus on the case
of decretion to demonstrate the absolute importance of long distance energy
transport, the order unity role played by the Poynting (or
by similar arguments the kinetic energy) flux means that
long distance energy transport also plays a crucial role in determining the
energy density of the fluctuating fields that drive Maxwell (or Reynolds) stresses in disks
accreting due to horizontal stresses powered by accretion energy.
Accordingly, the saturated strength of those fields everywhere but the innermost
edge of an accretion disk 
can be modulated by
 the rate at which energy is supplied by
the inner disk, but only on the time scale required for that modulation to change the stresses.
That time scale for the stresses to change sets a locality criterion: on longer time
scales the accretion rate in the outer disk depends on the inner disk.

\subsection{Inner Critical Radii and Disk Layering}

\begin{figure}[t!]\begin{center}
\includegraphics[angle=270, width=\columnwidth]{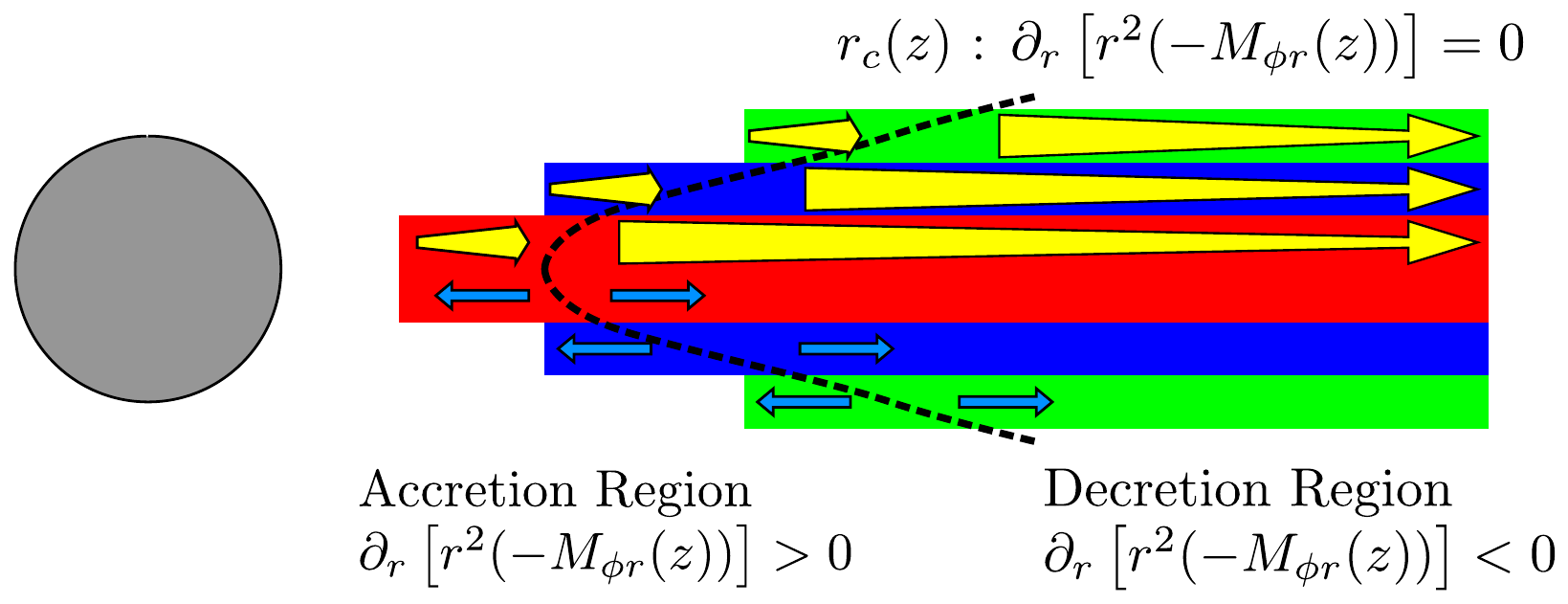}
\end{center}
\caption{
Structure of a disk dominated by horizontal Maxwell stresses.
In each layer,
decretion occurs beyond the height-dependent
cylindrical radius $r_c(z)$, where
the radial dependence of the Maxwell stress 
drops below
the critical value $\partial_r[r^2(- M_{\phi r}(z))] = 0$. 
Blue arrows in the bottom half show the direction of the accretion and decretion flow.
For horizontal stresses other than the Maxwell stress, the structure is the same, 
with the flux and stress replaced as appropriate.
Yellow arrows in the top half indicate the outward going Poynting flux, which 
grows though the accretion region, and deposits energy in the outer, decretion region,
although the transition between the growing and shrinking fluxes occurs at $9r_b(z)/4$,
which need not be the same as $r_c(z)$ and defines an alternate critical surface.}
\label{fig_fig1} 
\end{figure}

As we have reviewed above, in
steady state viscous disk models, the vertically integrated radial
mass flux is directed inwards while the midplane flow is directed
outwards.
This occurs because the much stronger radial gradients found at high altitudes drive
fast inward flows that overwhelm the midplane's slower outward  flows.
Horizontal stresses
drive horizontal energy fluxes (Sect.~\ref{sec_E_transport}).
Because the mass and energy fluxes are horizontal, this leads to a picture of
accretion disks as made up of layered, nearly horizontal slabs rather than adjacent annuli (see Figure~\ref{fig_fig1}).
In this picture, each individual layer at a height $z$ above the midplane has its own critical radius $r_c(z)$ that
defines the boundary between accretion and decretion.
In a vertically isothermal disk in hydrostatic equilibrium, the density is
\begin{equation}
\rho(r,z) = \rho_0(r) e^{-z^2/2H^2(r)},
\end{equation}
where $\rho_0$ is the midplane density and $H$ the scale height.  Taking the radial derivative
we find
\begin{equation}
 \partial_r \ln \rho(r,z) = \partial_r \ln \rho_0 +\frac{z^2}{H^2(r)} \partial_r \ln H(r).
\label{flaring_eq}
\end{equation}
The density, then, can only be taken to be a power-law in $r$ if the term capturing the radial dependence on $H$
is negligible compared to the radial dependence on the midplane
density, which requires $z^2/H^2 \ll 1$.
As noted in Section~\ref{sec_T_propto_p},
disk models with stresses strictly proportional to pressure
predict decretion where the disk can behave in a scale-free, power law manner, i.e.~where
the radial dependence on H in Equation~(\ref{flaring_eq}) can be neglected.
This implies that for each height $z$, the critical radius separating accretion and decretion will be near the radius
$r_{pl}(z)$ outside of which the disk, at the height $z$, can be approximated as a powerlaw in radius.
While the precise location of $r_{pl}$ will depend on the details of the disk, Equation~(\ref{flaring_eq}) gives the constraint
$H(r_{pl}(z)) > z$.

Further, standard viscous disk theory \citep{2002apa..book.....F} for vertically integrated disks 
invokes another critical radius $r_b$ at which $\partial_r \varOmega=0$.
Outside of 
\begin{equation}
r > \frac 94 r_b
\end{equation}
viscous accretion flows locally dissipate more energy than provided by the accretion power.
However, while vertical hydrostatic equilibrium can mathematically be extended to infinity, in practice
accretion disks will generally be truncated vertically by stellar or disk winds, or the presence of infalling envelope material.
It follows that each altitude $z$ has its own $r_b(z)$ where $\partial_r \varOmega(r,z)=0$.
Accordingly, a similar layered picture
results from a
analysis of viscous disks performed layer-by-layer, as opposed to the 
more common vertically-integrated approach.
 We can construct the surface
\begin{equation}
r(z)=\frac 94 r_b(z), \label{94r_star}
\end{equation}
noting that it divides an outer region centered on the midplane that
is a net consumer of energy from an inner, surface layer
that is a net exporter of energy.  
(Because disks generally 
increase in height with radius, their upper surfaces are also
their inner edges.)
The energetics of global disks depend fundamentally on
the material in the surface layer defined in Equation~(\ref{94r_star}), because in viscous disk theory
the accretion power below that surface at larger radii is inadequate
to power the dissipation \citep{2002apa..book.....F}.

\subsection{Available Accretion Energy}
\label{sec:power-decr-flows}

 Can accretion at the inner edge of each layer of the disk 
 provide enough energy to drive the outward motions expected further
out? 
The local surface density of the energy
released per unit time by accretion in a vertical slab of a Keplerian disk is
\begin{equation}
Q_{acc}= \int_z  \frac 12 \rho v_r r \varOmega^2 dz, \label{acc_power_0}
\end{equation}
where the factor of one half accounts for half of the gravitational
potential energy being converted into orbital kinetic energy. 
Azimuthally integrating Equation~(\ref{acc_power_0}) for a given slab,
we find the total power released at a radius $r$ to be 
\begin{eqnarray}
\nonumber \int_\phi Q_{acc} 2\pi r d \phi &= &
\frac 12 \left(2 \pi r \varSigma v_r r \varOmega^2\right)\\
& =& \frac 12 \dot{M} r \varOmega^2=\frac 12 \dot{M} r_f \varOmega_f^2 \left(\frac{r}{r_f}\right)^{-2}.
\label{acc_power_1}
\end{eqnarray}
where $\dot{M}$ is the accretion rate (so that $\dot{M} < 0$ implies
decretion), and $r_f$ is a fiducial reference position.  In Equation~(\ref{acc_power_1}) $\Sigma$ is the surface
density and $v_r$ is the density-weighted mean radial velocity of the slab.

Let us divide our slab into two annuli at $r_f$, which
we choose to be where $v_r=0$, such that there is an inner accretion
flow $\dot{M}_1>0$ with an inner edge at $r_0$, and an outer
outwards flow $\dot{M}_2<0$ out to $r_1$,
with $r_1 \gg r_f \gg r_0$. 
Then if the flows are to be
powered by the accretion, net gravitational energy must be
released by the system.  The accretion power provided by the inner annulus is
\begin{equation}
P_{in} = \int_{r_0}^{r_f} \frac {\dot{M}_1 r_f^3 \varOmega_f^2 r^{-2} }{2} dr= \frac 12 \dot{M}_1 \varOmega_f^2 \frac{r_f^2}{r_0} (r_f-r_0),
\end{equation}
while the accretion power provided by the outer annulus is (negative because $\dot{M}_2<0$)
\begin{equation}
P_{out} = \int_{r_f}^{r_1} \frac 12 \dot{M}_2 r_f^3 \varOmega_f^2 r^{-2} dr= \frac 12 \dot{M}_2 \varOmega_f^2 \frac{r_f^2}{r_1} (r_1-r_f).
\end{equation}
The condition that net gravitational potential energy be released by this flow, allowing for accretion-powered accretion stresses, is then
\begin{equation}
P_{in} + P_{out} >0,
\end{equation}
or
 \begin{equation}
 \left| \frac{\dot{M}_1}{\dot{M}_2} \right|> \frac{r_0}{r_1} \left(\frac{r_1-r_f}{r_f-r_0}\right) \simeq \frac {r_0}{r_f},
 \end{equation}
 where the inequality holds for the broad annuli we assume, with~$r_f/r_0 \gg 1$ and~$r_1/r_f \gg 1$.
Inner annuli accreting at a rate $\dot{M}_1$ therefore release adequate
 energy to drive broad regions of decretion with $|\dot{M}_2| \sim \dot{M}_1$, although this
 is only a necessary requirement, and the details of the energy transport also matter.

\subsection{Time Scales of Energy Transfer}
\label{subsec_time_scales}

The divergence of the Poynting (and analogous hydrodynamical energy) fluxes
is comparable to the accretion power and can drive broad outflowing regions.  
Therefore, we need to consider the impact that changes in the Poynting
flux, due, for example, to fluctuations or outbursts in the inner disk, 
will have on the stresses.
We can quantify this by following the analysis of 
BP99.
They derived the local steady state rate of energy dissipation
into heat (their Eq.\ 37) starting from 
the full energy evolution equation:
\begin{equation}
\frac{\partial \phantom{t}}{\partial t}\left(\frac 12 \rho v^2 + \rho \Phi+\frac{B^2}{2 \mu_0}\right) + \nab \cdot \FE = -\QQQ_e
\label{EQ_QQQ}
\end{equation}
(their Eq.\ 31).
To find the steady state rate, they assume that the time
derivatives are negligible, thus
identifying the divergence of the energy flux with the local
energy dissipation rate $\QQQ_e$ (positive $\QQQ_e$ means energy is
deposited as heat).
Note that the quantity we define as $\QQQ_e$ is the volume density
of the energy dissipation; so 
 the variable $Q_e$ defined by BP99,
the total energy dissipated
per unit time per unit surface area, is
\begin{equation}
Q_e = \int_z dz \QQQ_e.
\end{equation}
That is twice the total power that must be radiated from each disk
surface element.  Note that $Q_e$ is the surface power actually dissipated into heat
and radiated away, while
$Q_{acc}$ was the accretion power surface density.

In the case of accretion flows driven by Maxwell or turbulent Reynolds
stresses, the accretion or decretion power
is mediated by the energy density of the magnetic field or of the
turbulent flows.  If there is a mismatch 
between the local accretion power and the divergence of the energy fluxes, energy must be
injected into or extracted from the local magnetic or turbulent energy densities (see Equation~\ref{E1}
for the magnetic case).  For a slab, those
azimuthally averaged energy
surface densities are
\begin{equation}
\langle E_{T} \rangle_{\phi} = \int_z dz \left\langle \frac 12 \rho w^2 +\frac{B^2}{2\mu_0}\right\rangle_\phi,
\label{E_T_def}
\end{equation}
where we recall that the velocity has been decomposed into a mean orbital $\bar{v}$ and a fluctuating
$w$.

The terms that survive averaging are related to the stresses: we have $R_{ij} \equiv \rho w_i w_j$
and $M_{ij} \equiv \mu_0^{-1} B_i B_j$.  Accordingly we can write
\begin{equation}
E_T =  f \Sigma |W_{r \phi}|
\label{c_def}
\end{equation}
where $f>1$ measures the difference between the
  squared amplitudes of the magnetic and fluctuating velocity fields
 $B^2$ and $w^2$ and their horizontal correlators $|B_\phi B_r|$ and $|w_\phi w_r|$.
 The parameter $f$ relates the stresses to an energy density, and so is a relative of the Shakura-Sunyaev $\alpha$.
 Indeed, 
\cite{2011ApJ...738...84H} defined
 \begin{equation}
 \alpha_{mag} = -2 \frac{B_r B_\phi}{B^2},
 \end{equation}
finding $\alpha_{mag} \sim 0.3$.  We can therefore estimate $f^{-1} \sim \alpha_{mag}/2 \sim 0.1$ for MRI active disks.

 We can find the accretion power surface density in terms of the stresses from
Equation (28) of BP99,
\begin{equation}
P_{acc} = \Sigma R \varOmega_K^2 \langle u_r \rangle_\rho \sim 2 g \varOmega \Sigma W_{r\phi}
\end{equation}
where we have assumed that $\Sigma R^2 W_{r \phi} \propto r^{-g}$.  We see then that
changes in the divergence of the energy fluxes cause changes in
the local 
stresses and hence accretion rates on a time scale
\begin{equation}
\left| \frac{E_T}{P_{acc}}\right| =\left| \frac{f}{2g\varOmega} \right|.
\end{equation}
It immediately follows that local regions in accretion disks are only
buffered against 
changes in the incoming energy fluxes for orbital to tens of orbital time scales.
This means that the locality time scale for disk accretion
is only a few dynamical times, far shorter
than the global viscous evolution time.
This is most clearly the case in decreting regions, where the energy
that powers the decretion must come from somewhere, but even in accreting regions,
putting energy into or removing energy from the magnetic field or the turbulence will alter
the stresses.  Note that this time evolution of $E_T$ was set to zero in BP99
even when they considered the non-steady state to derive their
Equation 46, as we discuss in detail in Sect.~\ref{sec:BP99}.

\section{The Shearing Box Approximation}
\label{sec:shear-box-appr}

The shearing box approximation neglects
curvature terms and 
radial gradients of the hydrostatic disk background distribution of density and
temperature.
The curvature terms are consistently neglected by taking the horizontal size of 
the shearing box to be small compared to the radial position $r_0$.
This flattening of 
the differential operators allows replacement of
the cylindrical coordinate system with a local, Cartesian one.
To arrive at a system where
shear periodic boundary conditions can be applied in the radial direction, 
it is also necessary to 
neglect a disk's background radial density and temperature gradients. 
For the hydrostatic equilibrium
pressure $P_h$
that characterizes the global hydrostatic disk structure, this approximation about some radius $r_0$ is 
\begin{align}
P_h(r,z) = P_{h}(r_0,z) +  \mathcal{O}\left(\frac{r-r_0}{r_0}\right) \  \label{eq_phapprox}
\end{align}
\citep{1965MNRAS.130..125G,2004A&A...427..855U,2014arXiv1406.4864M}.
This approximation has significant consequences  
a few scale heights $H$
above the midplane, no matter how cold and thin the disk is,
because the approximation cannot capture a radially varying $H$.
In particular, meridional flows 
cannot be driven
once this approximation has been made.

The shearing box as commonly employed does, however, provide an internally consistent model for a shearing disk-like flow, 
if not a strictly valid asymptotic approximation to a section of an accretion disk in hydrostatic equilibrium. 
Given a shearing box, the underlying conservation properties of 
magnetohydrodynamics are preserved, up to specific effects of work done at the boundaries \citep{1995ApJ...440..742H,2014arXiv1406.4864M}.
As such, the shearing box has and can be used to understand the local dynamics in a shear flow closely analogous to that of a disk.
However, the mass flux through a vertical surface drawn in a disk cannot be determined
 from this shearing box model alone, due to the neglect of the radial derivatives 
of the background density and temperature structure and the resulting elimination of any meridional flows.

\subsection{The Poynting Flux in the Shearing Box}
\label{subsec_Energy_fluxes}

Consider an incompressible ideal unstratified shearing box.
where the velocity
$\bm{v}$ has been decomposed 
into $\barv+ \bm{w}$ with the background shear flow
$\bm{\bar{v}} = -q\varOmega x \ey$, $\bm{w}$ the fluctuation, and $q=3/2$ 
in the case of Keplerian rotation with $\varOmega \propto r^{-3/2}$.
The governing equations are:
\begin{align}
\partial_t \bm{w} &+ \barv\cdot \nabla \bm{w} + \bm{w}\cdot \nabla \bm{w} = \nonumber\\
& -2\varOmega \ez \times \bm{w} +q\varOmega w_x\ey 
   +\frac{1}{\mu_0 \rho} (\nabla\times\B) \times\B,  \\
\partial_t \B &= \nabla\times\left[ (\bm{w}+q\varOmega x\ey) \times \B \right],
\end{align}
with $\nabla\cdot \bm{w}=0$.

An exact solution of these governing equations is
\begin{align}
\bm{w}(t) &= 0, \\
\B(t) &=  \sin(k_z z) \left( \ex + q\varOmega t \ey \right), \label{eq_path_Bt} \\
\mu_0 \J(t) &= \nabla \times \B(t) = k_z \cos (k_z z) \left(-q \varOmega t \ex + \ey\right),
\end{align}
which has zero net magnetic field at all times.
The Lorentz force
\begin{equation}
\FF_L = \J \times \B = \frac{k_z}{\mu_0}  \sin(k_z z) \cos(k_z z) \left(1 +q^2 \varOmega^2\right) \ez
\end{equation} 
merely captures the magnetic pressure. 
The fluid is incompressible,
so no fluid is accelerated, the kinetic energy of the flow does not evolve and no work is done by
the flow.

The only time varying energy in the system is magnetic energy, which evolves according to:
\begin{align}
\frac{\partial}{\partial t} \left(\frac{B^2 }{2 \mu_0} \right)&= -\nabla \cdot \left[\frac{1}{\mu_0}\left(-\vv \times \BB +\eta\JJ \right)\times\BB\right] \nonumber\\
&\qquad -\eta \JJ^2 - (\JJ\times \BB) \cdot \vv.
\label{eq_Bener_evolution}
\end{align}
In the ideal limit ($\eta=0$), this has the solution
\begin{align}
\frac{B^2(t)}{2 \mu_0} & = \frac{\sin^2(k_z z) \left(1+ q^2 \varOmega^2 t^2\right) }{2 \mu_0} \ ,
\end{align}
which can also be obtained directly from Equation~\ref{eq_path_Bt}.

The rate of energy density growth is then
\begin{align}
\frac{\partial}{\partial t} \frac{B^2(t)}{2\mu_0} & = \frac{1}{\mu_0} \sin^2(k_z z) q^2 \varOmega^2 t \ . \label{eqpathsb_dtmagenergy}
\end{align}
No work is done inside the volume, so the energy must be provided by some flux from outside the volume, namely the
Poynting flux $\v{S}$.

For this exact solution, 
\begin{align}
\SSS &= \frac{1}{\mu_0}  q\varOmega x  \sin^2 (k_z z)  \left[  -  q \varOmega t \ex +  \ey \right]\, .
\label{eqpathsb_poynting}
\end{align}
The Poynting flux deposits energy into the field at the rate
\begin{align}
-\nabla \cdot \bm{S} &= \frac{ q^2 \varOmega^2  t}{\mu_0}  \sin^2(k_z z) \label{eqpathsb_divs}\ ,
\end{align}
which is the same rate that the 
magnetic energy grows (Equation~\ref{eqpathsb_dtmagenergy}).

We reemphasize that the shear does no work in this scenario, as the associated
Lorentz forces 
are directed in the $\ez$ direction, perpendicular to the fluid velocity in the $\ey$ direction, so
the fluid does 
not encounter any resistance to its movement. 
Therefore, even though the energy growth rate contains a factor of the shear rate $q\varOmega$ 
(Equation~\ref{eqpathsb_dtmagenergy}),
the role of the shear is  limited to
setting the field geometry, which yields the Poynting flux (Equations~\ref{eqpathsb_poynting} and~\ref{eqpathsb_divs}).

\subsection{Energy Fluxes in Shearing Boxes}

Starting with the  ideal MHD energy flux and 
generalizing BP99 (Eq.~32), one arrives at
\begin{align}
\FE = \bm{v} \left( \frac{1}{2} \rho v^2 + \rho \Phi + P \right) +\frac{\B}{\mu_0} \times (\bm{v} \times \B ) - \bm{\sigma'}(\bm{v}) \cdot \bm{v}
\end{align}
which includes the kinetic energy, potential energy, thermal energy, magnetic energy, and viscous fluxes.
    We use the same notation as in the previous section, but
in addition, $P$ is the thermal pressure and $\Phi$ is the gravitational potential.
The viscous stress tensor is
\begin{align}
\sigma'_{ik}(v) = \nu\rho \left(\frac{\partial v_i}{\partial x_k} +\frac{\partial v_k}{\partial x_i} -\frac{2}{3}\delta_{ik}\frac{\partial v_l}{\partial x_l}\right) + \zeta \delta_{ik} \frac{\partial v_l}{\partial x_l},
\end{align}
where $\nu$ is the viscosity, and $\zeta$ the bulk viscosity \citep{LandauAndLifshitz}.

In a shearing box, the $y$ and $z$ components of $\FE$ are periodic, but the $x$ component is not:
\begin{align}
\FE \cdot \ex
=
   & -q\varOmega x \rho  w_x w_y +w_x \left( \frac{1}{2} \rho (\bar{v}^2 +\bm{w}^2  )  + \rho \Phi + P \right) \nonumber \\
   &  +q\varOmega x \frac{1}{\mu_0} B_x B_y + \frac{1}{\mu_0} w_x B^2 - \frac{1}{\mu_0} B_x w_k B_k \nonumber\\
   &  +q\varOmega x \nu\rho \left[ \frac{\partial (-q\varOmega x)}{\partial x} \right]   +q\varOmega x \sigma'_{xy}(\bm{w})  \nonumber\\
   & -  w_y  \nu\rho \left[ \frac{\partial (-q\varOmega x)}{\partial x} \right]   -  w_i \sigma'_{x i} (\bm{w})
   \label{F_e_dot_e_x}
\end{align}
Hence, the divergence of the energy flux contains the terms 
\begin{align}
\nabla \cdot \FE
=
   & -q\varOmega  \rho  w_x w_y +q\varOmega  \frac{1}{\mu_0} B_x B_y -\nu\rho q^2 \varOmega^2  + ...
   \label{stress_terms}
\end{align}
This expression is essentially a differential version of equation~(8) from \citet{1995ApJ...440..742H}.
These are the energy sources that result from the non-periodic nature of the energy flux in the shearing box.
Note that in vacuum, kinetic energy fluxes are zero and Poynting fluxes reduce to radiation
fluxes.  This means that the energy
fluxes can have physically motivated boundary conditions when generalized from shearing boxes to global simulations
with disks embedded in vacuum.

\subsection{Viscosity-driven Energy Flux in the Shearing Box}

We can usefully compare shearing boxes behaving viscously due to
Maxwell, Reynolds (or even microphysical viscous) 
stresses to shearing boxes with imposed constant viscosity $\nu$.
A viscous shearing box flow with $\bm{w}=0$ also has a non-zero divergence of the energy flux \citep{1974MNRAS.168..603L}, 
but this is accompanied by work done against the viscous friction force.
To characterize this, it is useful to state the evolution equation for
kinetic energy in an incompressible viscous hydrodynamic flow
\begin{align}
\frac{\partial}{\partial t}\left( \frac{1}{2} \rho v^2\right) &= -\nabla \cdot \left[\rho \bm{v} \left(\frac{1}{2} v^2 +\frac{p}{\rho} \right) -\bm{v} \cdot \bm{\sigma'}\right] - \sigma'_{ik} \frac{\partial v_i}{\partial x_k} \ .
\label{eq_Kener_evolution}
\end{align}
In the steady $\bm{w}=0$ shearing box flow 
without Reynolds stresses,
this reduces to 
 a balance between the kinetic energy source from the
divergence of the kinetic energy flux, and the dissipation of kinetic energy to heat from the
 work done against  viscous friction:
\begin{align}
-\nabla \cdot \left[ -\bm{v} \cdot \bm{\sigma'}\right] &=\sigma'_{ik} \frac{\partial v_i}{\partial x_k} \ ,\\
-\nabla \cdot \left[ -\bm{\barv} \cdot \bm{\sigma'}(\bm{\barv}) \right]  &=  \sigma'_{yx}(\bm{\barv}) \frac{\partial \barv}{\partial x} \ , \\
 -\nu \rho q^2 \varOmega^2 &=  -\nu \rho q^2 \varOmega^2 \ .
\end{align}

The dynamics of the energy flux that results from constant
viscosity $\nu$ in a shearing box hence differ strongly from
that which results from the shearing of magnetic fields. This is because
the energy consumed by the shearing of magnetic fields can remain as magnetic energy,
increasing the Maxwell stresses and driving further accretion. 
By contrast, in the viscous case
the work done against viscous friction immediately thermalizes any energy left over from the 
radial energy flux, which includes the accretion or decretion power.
The assumption of constant viscosity means that this energy plays no further role in the accretion process.
This can be seen by contrasting the magnetic energy evolution (Eq.~\ref{eq_Bener_evolution})
to the kinetic energy evolution (Eq.~\ref{eq_Kener_evolution}), where in the kinetic energy case, 
the source of kinetic energy due to viscosity in the first term is exactly
balanced by the final term, which removes this energy to heat at the same time.

\section{Simulations of Accretion Disks}
\label{sec:simul-accr-disks}

\subsection{Applicability of Shearing Boxes}
\label{sec_interpretation_of_sb}

Shearing boxes may lead to unrealistic results when
used to model local regions in accretion disks controlled by
horizontal Maxwell stresses $M_{\phi r}$.  This occurs 
because the Poynting flux into the simulation domain
is unphysically set by the boundaries of the simulation
domain, with energy being supplied at the boundaries of
the periodic volume.
In a given domain, the energy in the magnetic field, and hence the magnetic stress itself,
depends on the energy supply.  Equivalent constraints also apply to horizontal Reynolds stresses
and their accompanying hydrodynamic energy fluxes.
Our conclusions about $M_{\phi r}$ are, however, primarily based on overall
energy considerations, and so do not address the extent to which local turbulent
properties can still be addressed by local models.

Other recent work
has indeed called that into question, though. One example is that the
MRI appears to expand to the artificially constrained box scale
\citep{2012MNRAS.422.2685S,2012ApJ...748...79Y,2014MNRAS.441.1855N}, again arguing 
that the local treatment must be 
carefully interpreted.
Another example is a comparison by \citet{2008A&A...481...21R} of
the growth of isolated MHD perturbations in the center of a periodic, 2D,
shearing box with the growth of the 
same perturbations in an otherwise identical
domain with free-slip, conducting walls on the radial boundary.
When the boxes had width $L_x = \pi$,
such that the radial boundaries were near the initial perturbations,
the shearing box generated spurious energy in the box. However, when
they moved the radial edges away by increasing the width of the box to 
$L_x = 2 \pi$, they found much better agreement between the shearing box and the
wall-bounded flow. Their interpretation of this phenomenon
was that the shearing box walls were adding energy to the
perturbations.  We interpret this as being due to the implied Poynting
flux (Sect.~\ref{subsec_Energy_fluxes}).

Nonetheless, there is
evidence that the small scale behavior of the MRI can be 
 modeled
in restricted domains.
\citet{2012ApJ...749..189S} examined the behavior of MRI in restricted
subdomains of a global model, and found that the correlations of the
turbulent fluctuations remained similar. 
Nonmodal analysis of the most energetic structures over finite growth times 
in a laboratory-like global MRI
configuration also suggests that the shearing waves 
are a dominant feature in that global problem, and are captured well in 
the local shearing box approximation \citep{2014arXiv1407.4742S}.

\subsection{ Existing Global Simulations}
\label{sec:glob-simul-exist}

In general, global simulations 
could show either
accretion or decretion locally at each radius.
Further, the boundary conditions can
force the flow.
\citet{2006A&A...457..343F} and \citet{2011A&A...534A.107F} find both accretion and decretion in 
global models, with accretion at small radii and decretion at larger radii.
Additionally, \citet{2011A&A...534A.107F} searched for and did not detect a meridional circulation
in an MRI active disk.  They found in this case
that the variation of stress is not proportional to the pressure in
the vertical direction
as would occur in a solution with vertically constant alpha, explaining their
lack of detection of the expected circulation.

\citet{2014ApJ...784..121S} performed global simulations with initial net vertical fields.
These simulations are especially interesting because they combine net vertical field with
a radial temperature gradient that forces vertical shear.
Their analysis of radial flows \citep[][Section~5.4]{2014ApJ...784..121S}
is consistent with our analysis.  Indeed they estimate that accretion
occurs in the interior section of their disk, and decretion occurs in
the exterior section as we predict. 

In these models, where some 
mass flows outwards and to higher total energy, the 
flow must be supplied with energy liberated from some other location where 
accretion occurs, or from the boundaries. Indeed, in each case, such locations exist,
consistent with our findings.

\subsection{Comparing Local and Global Simulations}

\cite{2010ApJ...712.1241S} compared the flux-stress relation from a global simulation with the results
for shearing boxes from \cite{2007ApJ...668L..51P}.
They found a qualitative agreement, but also a marked discrepancy, which they suggested
was due to inadequate resolution in the global simulations.  This resolution difficulty was also noted by
\cite{2011ApJ...738...84H} who attempted to derive resolution criteria from shearing boxes for use in global
simulations.  

From a different angle, \cite{2012ApJ...749..189S}
compared small sub-subdomains of a small subdomain of a global simulation with each other and with the original
subdomain 
   (but did not invoke the shearing sheet approximation).  
They found that there was general agreement between the sub-volumes,
confirming that local simulations 
   with physical boundary conditions
can be used to study small scale phenomena below the scale of their subdomain.  
Overall, our argument is consistent with published numerical simulations.

\subsection{Diagnostics}
\label{subsec-cons}

We have noted that shearing boxes are not subject to energy constraints present in accretion disks; and that
global simulations of accretion disks 
 may need to contend with locality time scales of a few tens of orbits.
This raises the
question of how meaningful our arguments are for global simulations:
while the locality time scale we find is just a few orbits, this only allows, but does not force, the disk to behave on
those time scales.  We propose two diagnostics to measure the effect in practice.

Firstly,
the outer-disk growth rate of magnetic instabilities such as the MRI will be partially slaved to the inner, faster
growing magnetic field. Radial fluxes originating in the rapidly evolving inner disk
should increase the outer disk's MRI growth rate beyond that following from a radially local analysis.
There are suggestions of this behavior in the literature
\citep{2010A&A...516A..26F}.  Secondly, the MRI has strongly time varying stresses 
\citep{1995ApJ...440..742H,2003ApJ...585..908F, 2008A&A...487....1B,2010ApJ...713...52D,
2011ApJ...738...84H,2011ApJ...735..122F, 2014ApJ...791...62M} and
we expect the magnetic stresses can be correlated across large radial extents
with a modest lag (on order of the orbital timescale).
There are tantalizing hints of such behavior: figure 5 of \cite{2010A&A...515A..70D} shows
the azimuthal magnetic field at $6$ AU being correlated with that at $4$ AU with a lag of only about
$100$ years, or about $10$ local orbits.  However, the correlations of
the stresses across radius and time are far less
clear (their figure 9).  Further numerical simulations
  with more focused diagnostics could better quantify these stress correlations.

\subsection{Energy Flux Boundary Conditions}
\label{sec_choice_BC}

If the Poynting flux at the inner edge of the simulation domain is not
zero, then it is either being imposed by the boundary conditions or set by the simulation volume itself.  
Unless the boundary condition for the Poynting flux is somehow
physically determined (e.g. by a stellar magnetosphere),
the energy made available to the system
will not be physically controlled, and may be significantly larger or smaller than the value in a simulation
that included the entire disk.
If, on the other hand, the Poynting flux is set to zero at a physical
location within
the accretion disk, then the energy
available to the simulation volume will be significantly below the
correct value because the accretion energy liberated
inside of this location
is not being transported any further 
into the domain.  This is the case for the common choice of
absorbing boundary conditions, which resistively destroy the magnetic field in buffer zones at the inner and outer edges of the
simulation domain, forcing the Poynting flux to zero.

In vacuum, the Poynting flux reduces to the radiative flux, justifying
radial boundaries positioned outside the disk with
zero Poynting flux boundary conditions.  At the outer edge this is reasonable: the evolution time scale
is long, and the outer edge of the simulation domain can be put far
from the disk, so that the outer edge of the disk 
will not cross the boundary during the lifetime of the simulation.
At the inner edge however, the edge of the computational domain 
cannot be put arbitrarily far from the disk edge, and the disk evolution time is fast, so the inner edge of the disk
will rapidly overrun the boundary.  Once that occurs, the appropriate outwards energy flux entering the simulation
domain from accreting gas cannot be tracked.

We need therefore to capture 
the structure of the inner edge of the disk directly.
This is possible in simulations that include the surface 
of the central object, such as black hole accretion simulations with horizon-penetrating coordinates
 and computational domains that extend inside the horizon \citep[ex:][]{2004ApJ...611..977M}.
For circumstellar systems where the accretion disc extends smoothly to the stellar surface, this surface
provides the appropriate location for the inner radial boundary \citep{2002MNRAS.330..895A,2002ApJ...571..413S}.
For circumstellar disks around magnetized stars, though, we suggest instead placing the inner boundary at the interface between the disk and
the stellar magnetosphere: energy released by matter accreting in the magnetosphere will travel along
the stellar magnetic field lines rather than entering the disk.
In practice, this means using a conducting porous boundary that allows both mass flow and magnetic flux detachment.
Note that this latter requires resolving the boundary layer that will form as the disk magnetic field piles up against
the stellar field, 
as is done for example by \citet{2013A&A...550A..99Z} in
two-dimensional models, and \citet{2012MNRAS.421...63R} in
three-dimensional models.  These models indeed show strong time-variability,
which our results suggest may be required to maintain accretion.

An alternative radial magnetic boundary condition may be to bracket
the simulation volume with dead zones dominated by non-ideal effects.
If the non-ideal effects are adequate to decouple an inner active
region of an accretion disk 
from the outer active region, then the Poynting flux will not be
dynamically important in the outer region.  However, in stratified disks with active surface layers,
this is unlikely to be the case, and recent results have shown that the Hall effect may allow large scale magnetic
fields to develop even in dead zones
\citep{2014A&A...566A..56L,2014ApJ...791..137B}.  Nonetheless,
conveniently, simulations using dead-zones as magnetic boundaries can
check their self-consistency post-hoc by evaluating the radial
Poynting flux in those self-same dead-zones, determining whether or not they
are dynamically significant.

\section{Extending BP99}
\label{sec:BP99}

In their classic work, BP99 
wrote the evolution of the non-thermal energy density
in a viscous disk in their Equation (31), reproduced below:
\begin{align}
\frac{\partial}{\partial t} &\left( \frac 12 \rho v^2 +  \rho \Phi +\frac{B^2}{8\pi}\right)  \nonumber \\
&+ \nab \cdot \left[\vv\left(\frac{1}{2}\rho v^2 + \rho \Phi +P\right) +\frac{\BB}{4\pi}\times(\vv\times\BB) \right] \nonumber \\
&=P \nab \cdot \vv - \eta_V \left(\delta_i v_j\right)\left(\delta_i v_j \right) -\frac{\eta_B}{4\pi} |\nab \times \BB|^2\, , 
\end{align}
where $\vv = R \varOmega \ephi +\uu$ is the complete velocity field, and 
$\varOmega$ is purely determined by the gravitational potential $\Phi$ (so $\uu$ includes terms deriving from radial pressure
support and does not have a zero average).  The square brackets contain the energy flux and
Gaussian CGS units are used. In this section we 
follow the notation of BP99  and use $R$ for the cylindrical radius.

When considering non-steady state cases (and after vertically integrating and averaging) they arrived at their Equation (46) for the energy
dissipated into heat per unit surface area per unit time, also reproduced below:
\begin{equation}
-Q_e =  \Sigma W_{R\phi} \frac{d\varOmega}{d\ln R}\, ; \label{Eq_46_base}
\end{equation}
see also our Section~\ref{subsec_time_scales} and Equation~(\ref{EQ_QQQ}).
In the analysis adopted by BP99, Equation~(\ref{Eq_46_base}) is accurate to leading order.
It is instructive to consider the full expression, with terms of all orders retained.
In deriving this expression we use the same procedure as in that work, with two exceptions.  
Firstly,  for a given quantity $X$ we average using
\begin{align}
\langle X\rangle_\rho &\equiv
   \frac{1}{2\pi R \Sigma}  \int_{-\infty}^{+\infty} \int_0^{2\pi}  \rho X R d\phi dz \, , \label{ouravg}\\
 \Sigma &\equiv   \frac{1}{2\pi R}  \int_{-\infty}^{+\infty} \int_0^{2\pi} \rho R d\phi dz\, .
\end{align}
Secondly we 
define a cylindrical part -spherical part decomposition of $\Phi$
\begin{align}
\Phi = \Phi_C(R) + \Phi_S(R,Z)\, , \quad \Phi_S(R,0) = 0\, .
\end{align}

After significant algebra using the simplifications laid out by BP99, but no 
approximations, we arrive at a version of Equation~(\ref{Eq_46_base}) with all terms retained:
\begin{align}
-Q_e =&  \Sigma   W_{R\phi}   \frac{\partial \varOmega }{\partial \ln R} + \AAA  + \BBB + \CCC + \DDD + \EEE \label{Eq_46_full}
\end{align}
\begin{align}
 \AAA =& \frac{\partial }{\partial t}  \Sigma \left\langle \frac{1}{2}u^2 + \frac{B^2}{8\pi\rho} \right\rangle_\rho  \\
                       =&   \Sigma \frac{\partial }{\partial t}   \left\langle \frac{1}{2}u^2 + \frac{B^2}{8\pi\rho} \right\rangle_\rho +
                           \left\langle \frac{1}{2}u^2 + \frac{B^2}{8\pi\rho} \right\rangle_\rho \frac{\partial }{\partial t}  \Sigma \nonumber \\
 \BBB =& \frac{\partial }{\partial t}  \Sigma R\varOmega \langle  u_\phi  \rangle_\rho \\
 \CCC =&- \frac{1}{R}\frac{\partial }{\partial R} R
       \Sigma \left\langle \frac{1}{2} u^2 u_R 
      + \frac{ B_z^2 }{4 \pi \rho} u_R  + \frac{  B_R B_z }{4 \pi \rho} u_z  \right. \nonumber \\
  &  \qquad \qquad \qquad \qquad \left. +\frac{B_\phi^2}{4 \pi \rho} u_R  
                            + \frac{B_R B_\phi}{4 \pi\rho} u_\phi   \right\rangle_\rho \\
  \DDD =& - \frac{1}{R}\frac{\partial }{\partial R} R \Sigma \left\langle \frac{ P}{\rho} u_R \right\rangle_\rho  \\
  \EEE =&   \frac{\partial }{\partial t} \Sigma \langle \Phi_S \rangle_\rho
    - \frac{1}{R}\frac{\partial }{\partial R} R \Sigma \langle \Phi_S u_R\rangle_\rho\, . 
\end{align}
This expression is exact for the averaging as specified by Equation~(\ref{ouravg}), however
it can also be taken as an approximation when expressed in terms of the mean flow variables by extending 
the averaging over an appropriate small radial interval $\Delta R$; 
these details are discussed in Appendix~\ref{appendix_averaging}.

It is difficult to determine the order at which small perturbed quantities enter the terms
on the right-hand side of Equation~(\ref{Eq_46_full}).
The perturbed velocity $u$ is no longer asymptotically small once turbulence has saturated;
we do not \emph{a priori} know the value of the ratio of perturbed velocity components
$\langle u_{\phi} u_r\rangle_\rho/\langle u^2\rangle_\rho$ but it
could easily be less than a tenth;  and, crucially, we do not know the time and length scales on which values vary.

However, we merely need to argue that effects beyond those captured in Equation~(\ref{Eq_46_base}) can arise.
To do that, we continue by adding only the first term $\AAA$ to Equation~(\ref{Eq_46_base}).  
We neglect the second term of $\AAA$, proportional to $\partial_t
\Sigma$, and $\BBB$, because we expect  the surface density, pressure
and hence also $\langle u_\phi \rangle_\rho$ to vary only slowly in
time, on the long, viscous time scale. 
We can neglect $\CCC$ and $\DDD$ in favor of $\AAA$ if background and fluctuating quantities 
vary more slowly in space than in time, i.e.~if the fluctuating terms
vary over length and time scales $\ell$ and $\tau$ such that
\begin{equation}
\frac{\ell}{\tau} \gg u.
\end{equation}
Finally, neglecting $\EEE$ implies that the disk is thin enough that
the spherical component of the potential $\Phi_S$ can be neglected.

Under those conditions, we arrive at an approximate form of Equation~(\ref{Eq_46_full}):
\begin{equation}
-Q_e \approx \Sigma   W_{R\phi}   \frac{\partial \varOmega }{\partial \ln R}
   +\Sigma \frac{\partial }{\partial t}   \left\langle \frac{1}{2}u^2
   + \frac{B^2}{8\pi\rho} \right\rangle_\rho.
\label{Eq_46_corr}
\end{equation}
We further recall from
Equation~(\ref{E_T_def}) that the energy in those reservoirs is related to the stress $W_{R\phi}$.  
Hence, the energy density in the rightmost term of Equation~(\ref{Eq_46_corr})
can be estimated as $f \Sigma W_{R\phi}$ by invoking Equation~(\ref{c_def}).
Neglecting the rightmost term of Equation~(\ref{Eq_46_corr}) therefore
amounts to neglecting the time derivative of the stresses.  
The term is however 
vital if considering physical processes such as the linear growth phase of MRI,
where the stresses grow on a timescale $\sim \varOmega^{-1}$.

It is further important because 
in Section~\ref{subsec_time_scales} we argued that the energy
contained in $E_T$ can only sustain the power $Q_e$ for a few orbits.
That means that a significant change in the incoming energy fluxes
must change $E_T$ on a time scale of only a few times $\varOmega^{-1}$. 
As a result, the two terms on the right-hand side of Equation~(\ref{Eq_46_corr}) are of the same order, and the rightmost term must be kept
if time variation in the stresses is contemplated.  
Future numerical experiments correlating stresses at different radii in global models
will be needed to move beyond these order-of magnitude estimates and
determine the actual speed of propagation of changes in the energy fluxes, and hence stresses.
To diagnose the energetics, the various terms of Equation~(\ref{Eq_46_full}) can be individually tracked in numerical simulations.

\section{Conclusions}
\label{sec_conclusions}

It has been recognized since the work of \cite{1974MNRAS.168..603L}
that large swathes of disks that accrete through horizontal stresses, 
including by mechanisms such as the MRI, do not
release adequate gravitational potential energy to power their dynamics.  This can be seen most clearly
in the context of meridional circulation and viscously spreading
disks: any parcel of gas flowing away from the central object cannot
be releasing gravitational potential energy.
 Indeed it is only in a narrow inner
annulus that the accretion power is greater than the dissipation of
energy into heat.  In regions where the accretion power is inadequate, 
the energy evolution equation must be 
balanced by long distance energy fluxes, such as Poynting flux traveling radially through
an MRI active disk, or stellar irradiation powering a thermal instability.
Further, the mismatch between the local
accretion power and energy dissipation is 
within an order of magnitude of the total
local energy density in the fields generating
the stresses, in the case of turbulent or magnetic stresses.  

We have described two important consequences for
accretion disks whose accretion stresses are powered by the accretion
energy itself, including MRI active disks.
Firstly, because the outer disk dissipates more energy than it provides, and imports the remainder
from the inner disk, the outer disk's behavior depends on the inner
disk's accretion rate.  It follows that
the energy flux,
produced on the fast, dynamical time scale of the inner disk,
 must be capable of modulating the stresses, and hence accretion rate
 even in the outer disk with its slower dynamical time scale.
In the case of global disks, this means that dynamics in the outer disk (which in viscous
disk models needs to import energy even though it is accreting, 
when vertically integrated)
should depend on the energy provided by the inner disk.  

In more practical terms, for numerical simulations this means that
energy fluxes through the radial boundaries 
can be important to first order 
in setting stresses and accretion rates.  
This behavior cannot be represented in shearing box simulations, where
the box sets its own boundary conditions and hence energy fluxes.  
Shearing boxes are therefore decoupled from
a constraint on the available energy that drives the 
magnetic and turbulent velocity fields exerting
stresses on the disk.
In the case where the energy fluxes play major roles, temporal changes in the fluxes should cause any stress-exerting turbulence
present to temporarily enter forced-turbulence or decaying-turbulence modes.  Even in steady states, the local power deficit means that
the forced-turbulence picture may apply.

Secondly, the energy fluxes are large compared with the local energy
densities associated with the accretion stresses in the sense that
the local energy density can only power horizontal accretion stresses for a few orbits.  If the time scale were much longer, for example
the global disk evolution time scale, then the
importance of the energy fluxes would be mitigated: the time scale for the outer disk to adjust itself to the energy fluxes
from the inner disk would be too long to matter.  However, because the time scale is short, the outer disk must
rapidly adjust to changes in the inner disk, which in turn has short dynamical time scales.
This short timescale may help explain
observations of time-variable accretion flows in protoplanetary disks
such as FUors and EXors \citep{2014arXiv1401.3368A}.

In sum, we have demonstrated that understanding and accurately treating long distance energy fluxes in disks
is vital in modeling and understanding accretion flows.  Changes 
in the energy fluxes in the outer disk due to changes in the accretion
rate in the inner disk impact the amplitude of the outer disk's stresses on
 dynamical, rather than viscous, time scales.
Additionally, we have suggested how to diagnose these phenomena in numerical simulations.
In particular, we suggest both
correlating stresses at different radii with time lags to determine the speed with which and the extent
to which the stresses in the outer disk depend on the stresses (and hence energy release and energy fluxes)
in the inner disk;
and tracking the full set of contributions to the local energy dissipation (the terms of Equation~\ref{Eq_46_full})
and comparing their measured magnitudes to empirically determine which are dominant.

\acknowledgements
We thank P.~Armitage, E.~Blackman, O.~Gressel, T.~Heinemann, G.~Lesur,
S.~Balbus
and many other colleagues,
including participants in the Facebook group Circumstellar Disks and Planet Formation, 
for comments that improved the manuscript.
The research leading to these results has received funding from the 
People Programme (Marie Curie Actions) of the 
European Union's Seventh Framework Programme 
(FP7/2007-2013) under REA grant agreement 327995 (CPM), and the U.S. National Science Foundation under 
CDI grant AST08-35734 and AAG grant AST10-09802 (WL, AH, JSO,
and M-MML).  AH received additional support from NASA OSS grant NNX14AJ56G.
 WL carried out his part of the work at the Jet Propulsion
Laboratory, under contract with the California Institute of
Technology, with support from the National Aeronautics and Space 
Administration (NASA) through the Sagan Fellowship Program
executed by the NASA Exoplanet Science Institute.

\bibliographystyle{yahapj}

\appendix
\section{Alternative Averaging}
\label{appendix_averaging}

To read Equation~(\ref{Eq_46_full}) as an approximation expressed in terms of the mean flow variables
one exchanges the averaging of Equation~(\ref{ouravg}) for
BP99's choice of averaging which includes averaging over an appropriate, small, radial interval $\Delta R$ given by:
\begin{align}
\langle X\rangle_\rho &\equiv
   \frac{1}{2\pi \Sigma \Delta R}\int_{-\infty}^{+\infty} \int_{R-\Delta R/2}^{R+\Delta R/2}\int_0^{2\pi} \rho X d\phi dR' dz\ , \label{BP99avg}\\
 \Sigma &\equiv   \frac{1}{2\pi \Delta R}\int_{-\infty}^{+\infty} \int_{R-\Delta R/2}^{R+\Delta R/2}\int_0^{2\pi} \rho d\phi dR' dz\, . \label{BP99sigma}
\end{align}
The averaging schemes are very similar, and the resulting expression Equation~(\ref{Eq_46_full}) we arrive 
at can be read using this averaging as approximation.
When replacing the averaging in Equation~(\ref{Eq_46_full}) with the form  Equation~(\ref{BP99avg})  the approximations implicitly 
invoked are of the form
\begin{align}
\int_{R-\Delta R/2}^{R+\Delta R/2} R' X dR' \approx R \int_{R-\Delta R/2}^{R+\Delta R/2} X dR' \, . \label{BP99DeltaRapprox}
\end{align}
Thus, the approximation is valid for $\Delta R/R \ll 1$.
 
\end{document}